\documentclass[prb,amsmath,amssymb,
 reprint]{revtex4-1}
\usepackage{graphicx}
\usepackage{dcolumn}
\usepackage{inputenc}
\usepackage{bm}
\usepackage{color}
\usepackage{graphicx}
\usepackage[english]{babel}
\usepackage{color}
\usepackage{booktabs,longtable}
\usepackage{natbib}
\usepackage{xspace}
\usepackage{textgreek} 
\usepackage{upgreek} 
\usepackage{mathrsfs}
\usepackage[colorlinks]{hyperref}
\hypersetup{
                pdfstartview={FitH},
                linkcolor=blue,
                citecolor=blue,
                filecolor=blue,
                urlcolor=blue
}

\newcommand{\fig}[2]{\textcolor{blue}{Fig.~\ref{Fig#1} #2)}}
\newcommand{\figsimple}[1]{\textcolor{blue}{Fig.~\ref{Fig#1}}}
\newcommand{\blue}[1]{\textcolor{blue}{#1)}}
\newcommand{\degreeC}{$^{\circ}$C\xspace}
\newcommand{\degree}{$^{\circ}$\xspace}

\newcommand{\tesla}{$~\textup{T}$\xspace}
\newcommand{\uA}{$~\text{\textmu A}$\xspace}
\newcommand{\um}{$~\text{\textmu m}$\xspace}

\newcommand{\copt}{(Co/Pt)$_n$\xspace}
\newcommand{\coptt}{(Co/Pt)$_2$\xspace}
\newcommand{\copttt}{(Co/Pt)$_3$\xspace}
\newcommand{\coptttt}{(Co/Pt)$_4$\xspace}
\newcommand{\vmcd}{$V_{\textup{MCD}}$\xspace}

\begin{document}

\begin{sloppypar}

\title{Electrical detection of magnetic circular dichroism: application to magnetic microscopy in ultra-thin ferromagnetic films}

\author{T. Guillet}
\affiliation{Univ. Grenoble Alpes, CEA, CNRS, Grenoble INP, IRIG-SPINTEC, 38000 Grenoble, France}
\author{A. Marty}
\affiliation{Univ. Grenoble Alpes, CEA, CNRS, Grenoble INP, IRIG-SPINTEC, 38000 Grenoble, France}
\author{C. Vergnaud}
\affiliation{Univ. Grenoble Alpes, CEA, CNRS, Grenoble INP, IRIG-SPINTEC, 38000 Grenoble, France}
\author{F. Bonell}
\affiliation{Univ. Grenoble Alpes, CEA, CNRS, Grenoble INP, IRIG-SPINTEC, 38000 Grenoble, France}
\author{M. Jamet}
\affiliation{Univ. Grenoble Alpes, CEA, CNRS, Grenoble INP, IRIG-SPINTEC, 38000 Grenoble, France}

\date{\today}

\begin{abstract}

Imaging the magnetic configuration of thin-films has been a long-standing area of research. Since a few years, the emergence of two-dimensional ferromagnetic materials calls for innovation in the field of magnetic imaging. As the magnetic moments are extremely small, standard techniques like SQUID, torque magnetometry, magnetic force microscopy and Kerr effect microscopy are challenging and often lead to the detection of parasitic magnetic contributions or spurious effects.
In this work, we report a new magnetic microscopy technique based on the combination of magnetic circular dichroism and Seebeck effect in semiconductor/ferromagnet bilayers. We implement this method with perpendicularly magnetized (Co/Pt) multilayers sputtered on Ge (111). We further show that the electrical detection of MCD is more sensitive than the Kerr magnetometry, especially in the ultra-thin film regime, which makes it particularly promising for the study of emergent two-dimensional ferromagnetic materials.
\end{abstract}

\maketitle

\section{Introduction} 


With the recent emergence of two-dimensional ferromagnets \cite{CrI3,CGT,FGT}, magnetic imaging techniques have to be pushed to their ultimate detection limits to sense very low magnetic moments and stray fields. In this respect, several advanced scanning magnetic probe microscopies have been successfully used to image the magnetic configuration of ultra-thin ferromagnets down to the monolayer limit. For instance, magnetic force microscopy (MFM) and NV-center microscopy are sensitive to the magnetic stray field from the film by using a magnetic tip and a single NV spin in diamond respectively\cite{MFM,NV}. Spin-polarized scanning tunneling microscopy (SP-STM) is probing the unbalance between spin up and down densities at the Fermi level by tunneling magnetoresistance between the magnetic material and the atomically sharp magnetic tip. It could be used to image magnetic domains in Fe$_3$GeTe$_2$ at low temperature\cite{SPSTM}. Electron microscopies like transmission electron microscopy in the Lorentz mode or scanning electron microscopy with polarization analyzer (SEMPA) were also used to image the magnetic domains and skyrmions in Fe$_3$GeTe$_2$\cite{Lorentz,SEMPA}. Both rely on the interaction between electrons and the magnetic film. Magneto-optical Kerr effect (MOKE) and photoemission electron microscopy combined with x-ray magnetic circular dichroism (XMCD-PEEM) relying on the light-matter interaction were used either in scanning mode or in far field to probe the magnetic domains in several 2D ferromagnets\cite{MOKE,XMCD}. Finally, the intrinsic semiconducting properties of the ferromagnet itself could be used to image magnetic domains in CrBr$_3$ taking advantage of the optical selection rules for the absorption and emission of circularly polarized light\cite{PL}. Ultra-thin films are almost transparent for light and it is possible to use the substrate on which the material was grown or transfered to perform magnetic imaging. Indeed, the transmitted light interacts with the ferromagnet through magnetic circular dichroism and can be analyzed electrically by using the thermoelectric or photoelectric effects in the semiconducting substrate. If the substrate exhibits strong photoresponse, this last hybrid technique combining light and electrical measurements can be very sensitive to the magnetic state of the ultra-thin ferromagnet. By scanning the light beam, the magnetic configuration can be easily mapped at a submicrometer scale with high signal-to-noise ratio.


In this work, we report the growth of ultra-thin perpendicularly magnetized electrodes on Ge (111). Germanium exhibits strong thermoelectric and photoelectric responses\cite{Ioffe}. In order to obtain perpendicular magnetic anisotropy (PMA), we grow (Co/Pt) multilayers thin films using magnetron sputtering. In these films, the reduced symmetry and spin-orbit coupling at the interface between Co and Pt are responsible for the PMA \cite{daalderop_prediction_1992,johnson_orientational_1992}. Moreover, it was shown that the PMA increases with the number of repetitions (i.e. the number of interfaces).\cite{yakushiji_ultrathin_2010,grenet_origin_2011}. We probe the local magnetization orientation using simultaneously the anomalous Hall effect, Kerr microscopy and a new original technique based on the helicity dependence of the photovoltage in Ge by the magnetic circular dichroism (MCD) in (Co/Pt). This technique relies on the Seebeck effect in Ge. We study the (Co/Pt) thickness dependence of the Kerr effect and the MCD signal by changing the number of (Co/Pt) repetitions and we demonstrate that this MCD-based detection becomes much more sensitive than the Kerr effect in the ultra-thin film regime, which is promising for the future investigation of the magnetic properties of two-dimensional ferromagnets.
	
	\section{Sample preparation and experimental setup}
	
	
In this study, we use a 2\um- thick Ge/Si (111) film deposited by low-energy plasma-enhanced chemical vapor deposition (LEPECVD)\cite{gatti_gesige_2014}. The deposition rate was $\approx$4 nm s$^{-1}$ and the substrate temperature was fixed at 500\degreeC. Post-growth annealing cycles have been used to improve the crystal quality. The Ge layer is non intentionally doped with a residual electron carrier concentration $n$ $\approx 2 \times10^{16} \textup{cm}^{-3}$ as measured by Hall effect at room temperature.

This low $n$-doped 2 \um -thick Ge/Si (111) substrate is subsequently cleaned in acetone and isopropanol in an ultrasonic bath for 5 minutes to remove organic species. Then the substrate is dipped into a 50 \% hydrofluoric acid solution to remove the native Ge oxide and is transferred to the sputtering chamber. We do not heat the substrate during the growth as it promotes the chemical reaction between Co and Ge atoms at the interface, which is detrimental for the magnetic properties. 	
The chamber base pressure is in the 10$^{-8}$ mbar range. After introducing the sample, we set the Ar pressure in the chamber to $P_{Ar} \approx 1.2\times 10^{-2}$~mbar using a flowmeter. A 5 W DC power is applied to generate the plasma, giving a deposition rate of 0.25 \AA /s for Co and 0.79 \AA /s for Pt as measured by a quartz microbalance. We start with the deposition of a 0.5 nm-thick Co layer and end with a 1.8 nm-thick Pt layer which also acts as a capping layer preventing Co oxidation under atmospheric conditions. In this work, we grew \copt samples where the (Co/Pt) bilayer is repeated from one to four times ($n=1,2,3$ and $4$).

\begin{figure}[h!]
\begin{center}
\includegraphics[width=0.5\textwidth]{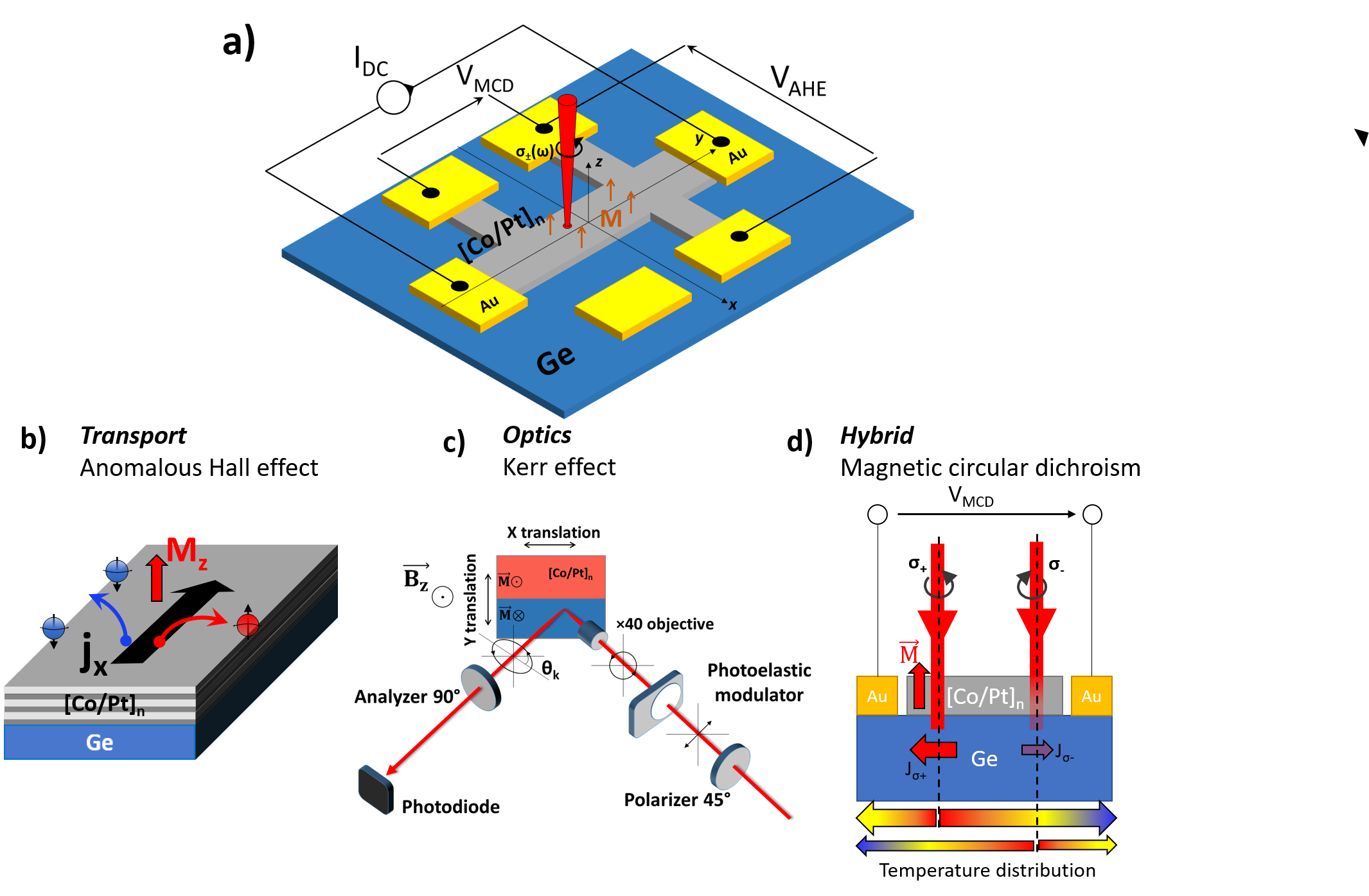} 
\caption{(color online) a) Sketch of the \copt /Ge (111) sample layout used for magnetic microscopy experiments. b) Anomalous Hall effect c) Magneto-optical Kerr microscopy d) Helicity dependent photovoltage due to MCD: the difference of transmitted power between $\sigma_{+}$ and $\sigma_{-}$ light helicities results in a difference of temperature distribution in the sample. This difference is recorded electrically using the Seebeck effect of Ge: the magnetic configuration first translates into a thermal information and then into an electrical one.}
\label{Fig1}
\end{center}
\end{figure}


We then proceed with the definition of $200\times 50~$\textmu m$^2$ Hall bars in the \copt film. We first use the laser lithography technique to define the conduction channel and we etch the \copt film using ion beam etching (IBE). Electrical contacts are lithographically defined and Ti(5 nm)/Au(120 nm) contacts are deposited by e-beam evaporation. The final device is sketched in \fig{1}{a}. The electrical contacts allow for magnetic characterizations by magnetotransport measurements and the channel is large enough to perform Kerr microscopy. One contact is not connected to the Hall bar, in order to measure the voltage between the ferromagnetic film and the Ge substrate and to detect a possible non-local spin signal. 


As shown in \fig{1}{b}, a DC current $I_{\textup{DC}}$ is applied in the (Co/Pt) bar, the electrons are deflected transversely as a consequence of the anomalous Hall effect (AHE). The transverse resistance, defined as $R_{AHE}$, is proportional to $M_z$, the out-of-plane component of the magnetization. 

In the meantime, we perform magneto-optical Kerr effect (MOKE) imaging of the magnetization. The sample is illuminated with a circularly polarized laser beam, the circular polarization ($\sigma_{\pm}$) is modulated at $f = 42$ kHz by using a photoelastic modulator (PEM). The reflected light is then analyzed by a polarizer and the light intensity is recorded using a photodiode. The resulting photovoltage is demodulated at $2\omega$ by a lock-in amplifier, to obtain the Kerr rotation $\theta_k$ (see \fig{1}{c}). \newline

The (Co/Pt) film being very thin, the circularly polarized light is partially transmitted through the film and electron-hole pairs are photogenerated in Ge. As the (Co/Pt) magnetization is perpendicular, the left and right circularly polarized photons have different transmission coefficients due to MCD \cite{stephens_theory_1970} (see \fig{1}{d}). 
Due to light absorption, the Ge layer is locally heated at the position of the laser spot and a Seebeck voltage $V_{\textup{DC}}^{\textup{Seebeck}}$ develops between the Au electrodes in \fig{1}{d}: $V_{\textup{DC}}^{\textup{Seebeck}}=V_{\sigma_+} = S \Delta T_{\sigma_+}$ for the $\sigma_+$ polarized light and $V_{\textup{DC}}^{\textup{Seebeck}}=V_{\sigma_-} = S \Delta T_{\sigma_-}$ for the $\sigma_-$ polarized light, $S$ being the Seebeck coefficient of Ge and $\Delta T_{\sigma_+}$ (resp. $\Delta T_{\sigma_-}$) the temperature difference between the Au electrodes for the $\sigma_+$ (resp. $\sigma_-$) polarized light. Note that if the laser spot is exactly located in the middle of the two Au electrodes, the Seebeck voltage is zero for both helicities. 
Since the $\sigma_+$ and $\sigma_-$ polarized lights are differently absorbed in Ge due to the MCD in the (Co/Pt) layer, $V_{\sigma_+} \neq V_{\sigma_-}$ and we detect a voltage \vmcd at the PEM frequency as a combination of the Seebeck effect in Ge and the MCD in (Co/Pt).

The DC and demodulated voltages $V_{\textup{DC}}$ and \vmcd are simultaneously recorded with a nanovoltmer and lock-in amplifier, respectively, while the magnetic field is swept. Alternatively, we can fix the magnetic field and image the sample magnetic configuration by scanning the laser beam at normal incidence. \newline

	\section{Anomalous Hall effect, Kerr effect and electrical detection of the MCD}	


We first focus on the \copttt sample, \fig{2}{a} shows the sample reflectivity recorded by scanning the laser beam on the microstructure. The (Co/Pt) Hall bar pattern is in green, the Au/Ti contacts in red and the Ge substrate in blue. The circularly polarized laser beam is first focused on the center of the Hall bar (at the position of the red spot). \fig{2}{b-d} show the magnetic signals for a $\pm 500$~Oe magnetic field sweep, applied perpendicularly to the film plane, recorded simultaneously using the three aforementioned techniques. All the measurements are performed at room temperature.  

\begin{figure}[h!]
\begin{center}
\includegraphics[width=0.5\textwidth]{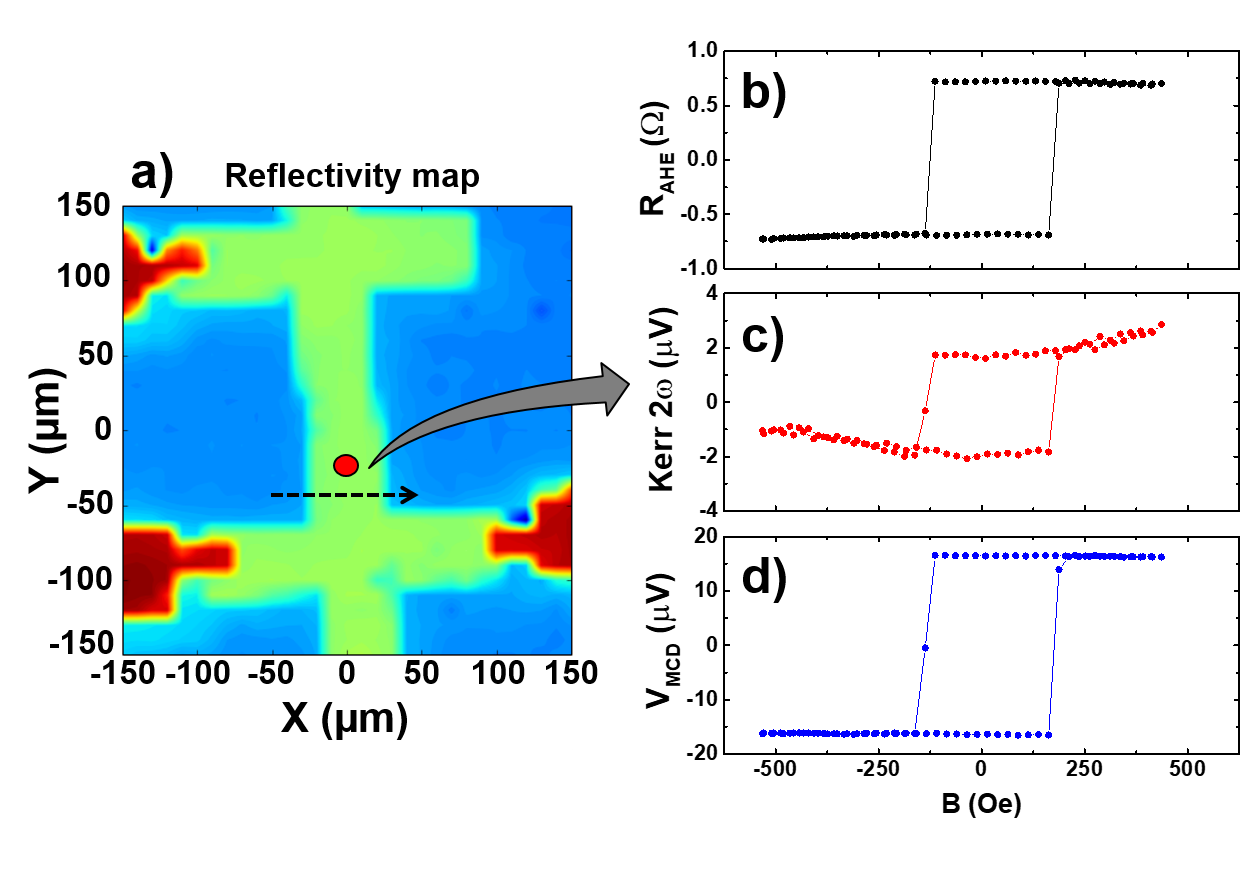} 
\caption{(color online) a) Two-dimensional reflectivity map of the \copttt /Ge Hall bar, the red circle indicates the laser beam position during the magnetic field sweep (applied perpendicularly to the sample plane). The black dashed line corresponds to the line scan along $x$ of \fig{3}{a-d}. b) AHE hysteresis loop . c) MOKE hysteresis loop using a 100 \% circularly polarized red light ($\lambda = 661$~nm) focused on the Hall bar center, the spot size is about 1.5\um. d) \vmcd hysteresis loop. The voltage is demodulated at the PEM frequency $\omega$ and is measured between a Hall bar contact and the substrate, a current $I_{\textup{DC}} = 100$\uA is applied during the measurement.}
\label{Fig2}
\end{center}
\end{figure}

In this geometry, the observation of a square hysteresis loop indicates that the \copttt sample magnetization is out-of-plane. For this $n=3$ repetitions sample, the coercive field is $B_c \approx 160$~Oe. We also note that the MCD signal is one order of magnitude larger than the Kerr signal, so the technique looks interesting for ultra-thin ferromagnetic films where the Kerr signal amplitude is very small.

\begin{figure}[h!]
\begin{center}
\includegraphics[width=0.5\textwidth]{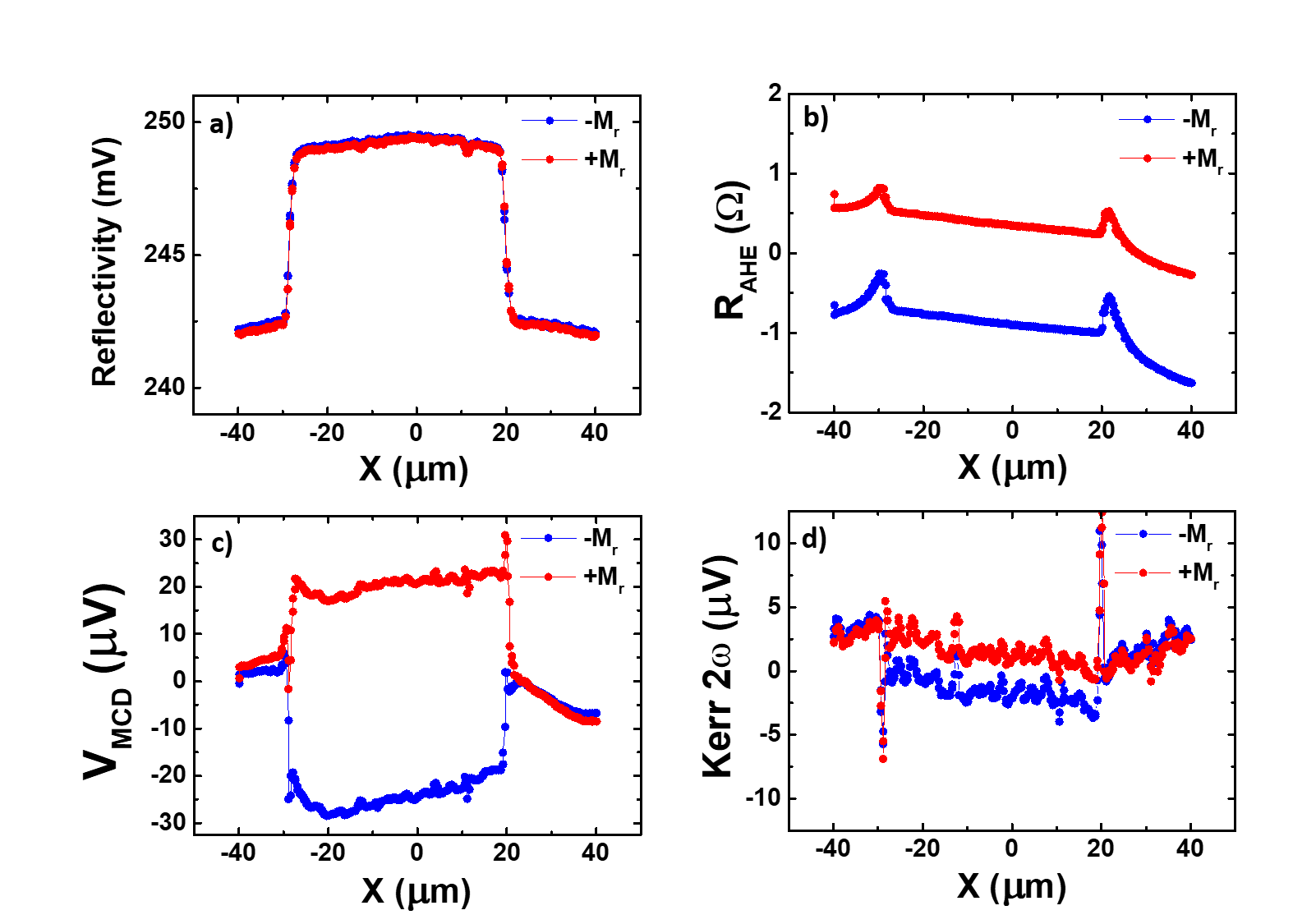} 
\caption{(color online) Line scans along the $x$ direction (black dashed line in \fig{2} {a}) of the remanent magnetic states $+M_r$ along $+z$ in blue and $-M_r$ along $-z$ in red ($B = 0$\tesla). a) Sample reflectivity. b) Anomalous Hall effect c) \vmcd and d) Kerr angle.}
\label{Fig3}
\end{center}
\end{figure}

The anomalous Hall effect gives a macroscopic picture of the magnetization, whereas the MOKE and MCD techniques can be spatially resolved by scanning the sample with the laser beam. We perform line scans along the $x$ direction. We first apply $+500$~Oe or $-500$~Oe along $z$ to saturate the film magnetization either up or down and then record the corresponding remanent state $+M_r$ or $-M_r$ at zero field. \fig{3}{a} shows the sample reflectivity, the (Co/Pt) film being more reflective than Ge, it corresponds to the central area where the photodiode signal is the largest.
\fig{3}{b} reports the AHE line scans, a weak spatial dependence of the signal is observed as a consequence of the Seebeck effect that takes place due to the scanning laser spot heating locally the Ge film (this contribution can be removed by using an AC current and a lock-in detection to measure the AHE). 
\fig{3}{c} and \blue{d} show the remanent magnetization measured by \vmcd and Kerr effect, respectively. A clear contrast can be observed in both cases and we confirm the local nature of the MCD signal: when the laser beam directly illuminates the Ge film, the \vmcd signal vanishes. Again, we note that the \vmcd signal is more than one order of magnitude larger than the Kerr effect signal. In order to better understand the nature of the \vmcd signals, we then performed large two-dimensional maps of the magnetic configuration.
	
The magnetization is first initialized in the $+M_r$ remanent state by applying a $+500$~Oe external magnetic field along $+z$. \fig{4}{a} shows the sample reflectivity, the Hall bar contours are highlighted by a black dashed line. \fig{4}{b} and \blue{d} show the \vmcd signal and the DC photovoltage $V_{\textup{DC}}$, respectively, using the contacts configuration shown in \figsimple{1}{a}. 
We observe that the DC photovoltage is positive when the laser beam scans the top area ($ Y > 0$\um) and negative in the bottom area ($ Y < 0$\um). It corresponds to the Seebeck voltage in Ge due to the temperature difference between the two electrical contacts induced by the laser spot heating. Interestingly, we observe the same behavior for the \vmcd signal (demodulated at the PEM frequency). By using both the DC and MCD photovoltages, we can first calculate $\gamma$, the MCD signal (in \%) of the (Co/Pt) film: $\gamma = V_{\textup{MCD}}/V_{\textup{DC}} \approx 0.3 \%$. 
This normalization can be performed point-by-point, for each position of the laser beam and results in a position-independent map of the magnetic configuration. The DC photovoltage intensity also allows us to estimate the temperature gradient in the Ge channel using the Seebeck effect relation and the Seebeck coefficient of Ge ($S = 330~$\textmu V/K)\cite{Ioffe}. $\Delta T = V_{\textup{DC}}^{\textup{Max}}/S \approx 36$ K .  \newline

\begin{figure}[h!]
\begin{center}
\includegraphics[width=0.5\textwidth]{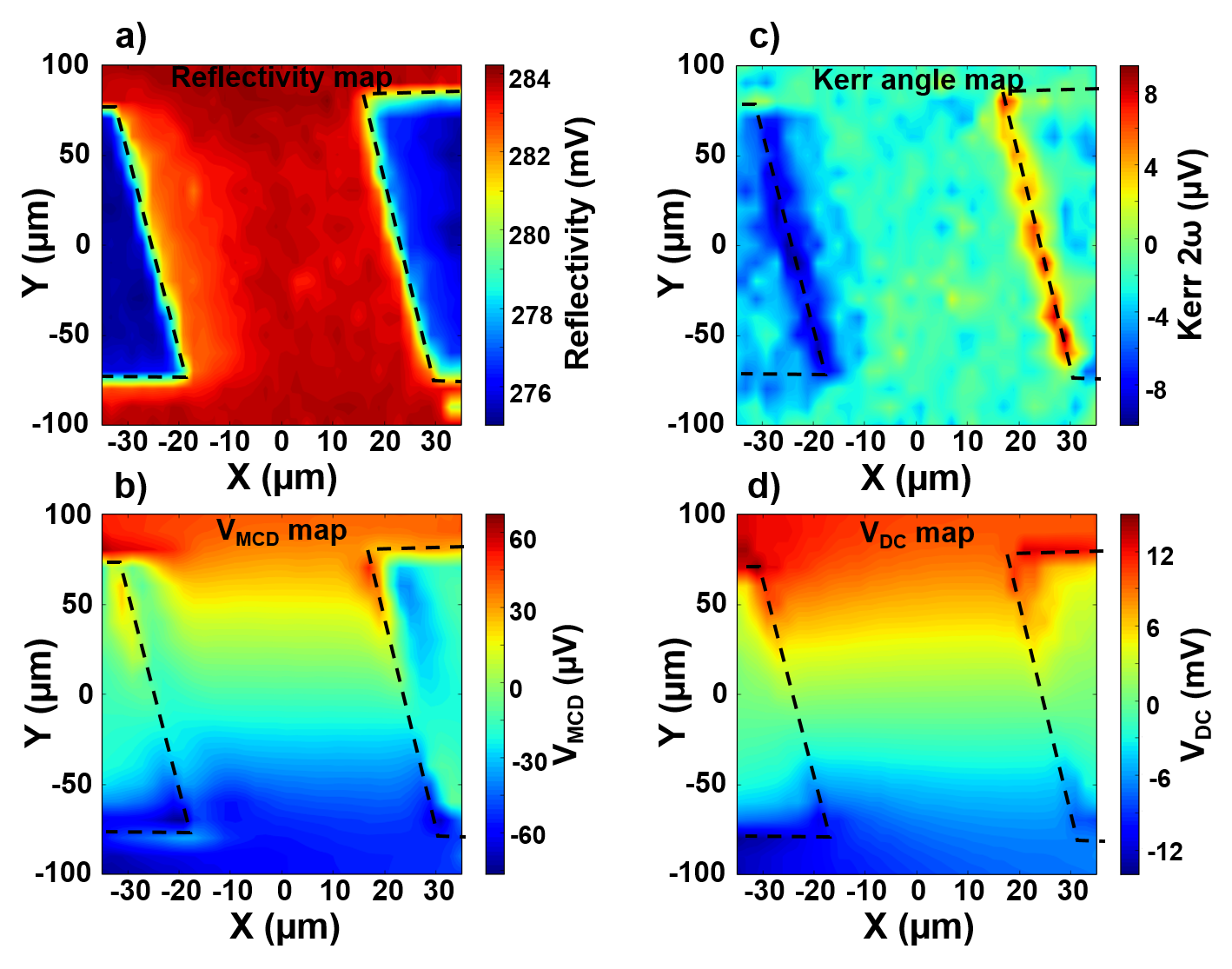} 
\caption{(color online) Two-dimensional maps of the remanent magnetic states $+M_r$ along $+z$ ($B = 0$\tesla) for $I_{DC}=0$ A. a) Sample reflectivity. b) \vmcd c) Kerr angle and d) DC photovoltage.}
\label{Fig4}
\end{center}
\end{figure}


To further understand how the \vmcd signal is affected by the temperature distribution in Ge when scanning the laser beam, we record hysteresis loops for different vertical positions ($Y$) of the laser spot on the Hall bar. 
\fig{5}{a} shows that the hysteresis loop signal is reversed between $ Y > 0$\um and $ Y < 0$\um while the Kerr effect is independent of the beam position (\fig{5}{b}). The difference of signal between the two remanent states is plotted as a function of $Y$ in \fig{5}{c}, we clearly see the \vmcd signal changing with the laser beam position. \newline

\begin{figure}[h!]
\begin{center}
\includegraphics[width=0.5\textwidth]{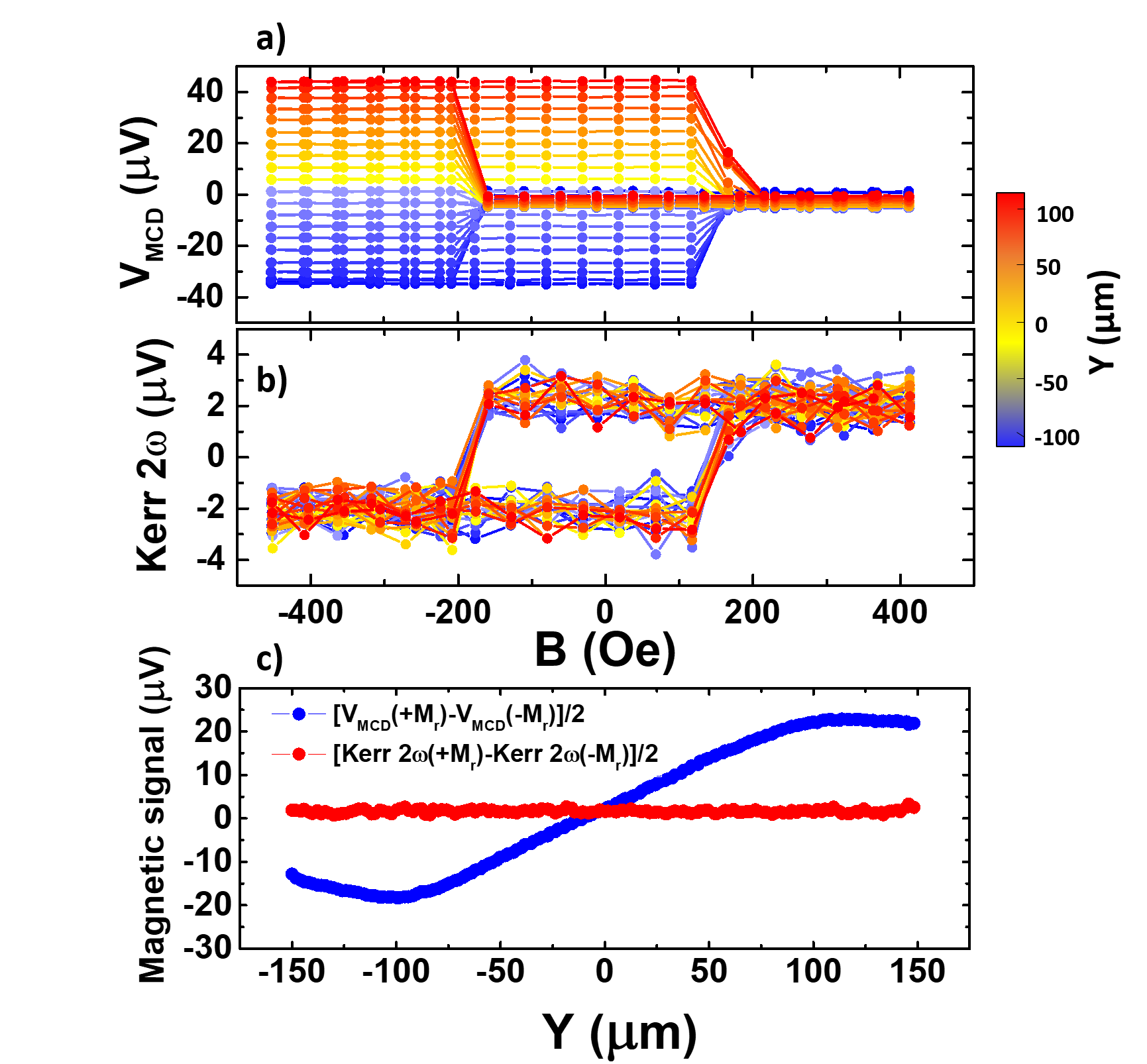} 
\caption{(color online) a) \vmcd and b) Kerr angle hysteresis loops recorded for different vertical positions of the laser beam on the Hall bar. c) Difference of signal between the positive and negative remanent states from \vmcd and Kerr angle as a function of the position of the beam on the Hall bar.}
\label{Fig5}
\end{center}
\end{figure}


The \vmcd signal being geometry-dependent, it is not suitable and reliable to perform magnetic imaging. Several approaches can be used to solve this problem. First, one can simply normalize the \vmcd signal by the DC photovoltage $V_{\textup{DC}}$ to obtain an almost position-independent measurement. However, in the region located in the middle of the two electrical contacts, the sensitivity of this technique vanishes.

One can optimize the contacts geometry to have an almost uniform temperature in Ge at the level of the magnetic microstructure to image regardless of the laser beam position by patterning one contact close to the microstructure and a second one far away. This would optimize the Seebeck effect-based detection of the magnetic circular dichroism. Moreover, using a material with a large Seebeck coefficient like Ge ($S = 330~$\textmu V/K) is necessary to obtain large signals. 

An alternative technique consists in applying a bias current through the Hall bar, along the MCD electrical detection axis (along $y$ here). In this way, the charge carriers photogenerated by the laser beam are drifting along the applied bias electric field. However, due to MCD, the densities of photogenerated charge carriers for $\sigma_+$ and $\sigma_-$ polarized lights are different giving rise to a modulated voltage $V_{\textup{MCD}}^{\textup{drift}}$ at the PEM frequency. Then, the total $V_{MCD}$ signal contains both the Seebeck MCD voltage $V_{\textup{MCD}}^{\textup{Seebeck}}$ and this drift component $V_{\textup{MCD}}^{\textup{drift}}$ except that the Seebeck voltage is even (independent) with respect to the bias current direction while the drift component is odd. We report this type of measurement in \figsimple{6}. The magnetization is prepared in $-M_r$ state by applying a $-500$~Oe  external magnetic field. A DC bias current is dynamically applied between the two detection contacts and the even and odd components of the \vmcd signal with respect to the current are calculated and plotted as a function of the position of the laser spot. \fig{6}{a} shows two-dimensional maps of the bias current-dependent voltage $V_{\textup{MCD}}^{\textup{odd}}$ for bias currents from 20\uA to 100\uA , the corresponding profiles for $X = 0$\um are shown in \fig{6}{c}. We observe a clear spatial-independent \vmcd signal, which varies linearly with the bias current. The current-independent component $V_{\textup{MCD}}^{\textup{even}}$ maps are reported in \fig{6}{b} and the corresponding profile for $X = 0$\um in \fig{6}{d}, we find again the fingerprint of the Seebeck effect-based MCD detection.\newline
 
\begin{figure}[h!]
\begin{center}
\includegraphics[width=0.5\textwidth]{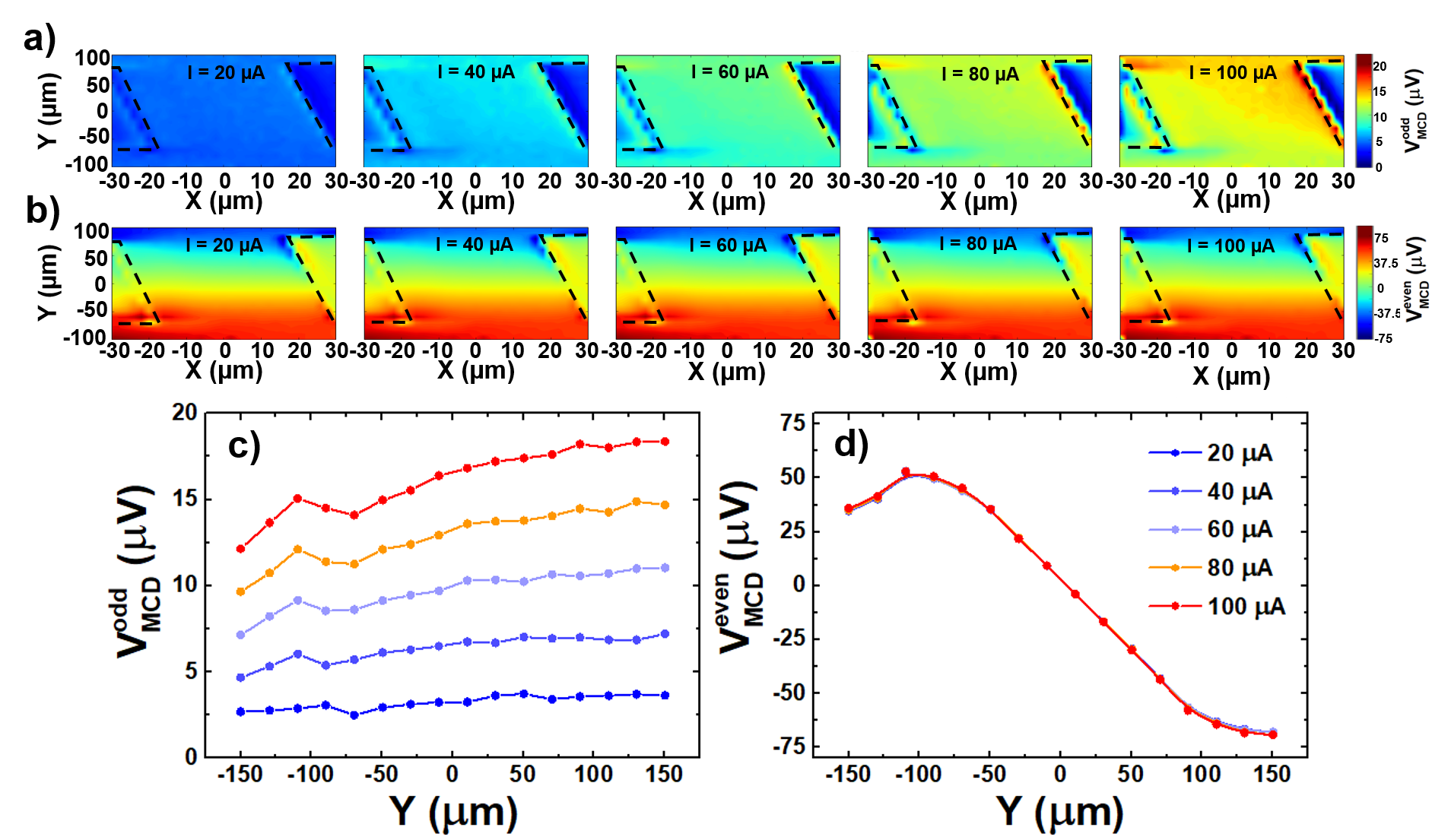} 
\caption{(color online) a) Two-dimensional maps of the bias dependent (odd with $I$) component of the \vmcd signal for bias currents from 20\uA to 100\uA. b) Corresponding bias independent (even with $I$) component. c) $V_{\textup{MCD}}^{\textup{odd}}$ profiles at $X = 0$\um. d) $V_{\textup{MCD}}^{\textup{even}}$ profiles at $X = 0$\um.}
\label{Fig6}
\end{center}
\end{figure}


In the following, we do not apply any bias current and take advantage of the position of the disconnected contact far from the the Hall bar to maximize the Seebeck effect-based detection of MCD. By using this configuration, the scanning area is far from the middle of the two detection contacts and the Seebeck voltage (\textit{i.e.} \vmcd signal) is almost independent of the position of the laser beam on the scanned area. This is necessary to have a reliable magnetic image of the (Co/Pt) microstructure. We first investigate the dependence of the magnetic signals as a function of the repetition number $n$ of (Co/Pt) bilayers. \figsimple{7} summarizes the results where the magnetization is measured simultaneously using the \vmcd and the Kerr effect. The light beam is focused on the center of each Hall bars as illustrated in \fig{1}{a}. When sweeping the magnetic field perpendicularly to the film plane, hysteresis loops are observed, indicating that all the films show PMA. We can also notice that the coercive field increases with the number of repetitions, as a consequence of a larger magnetic anisotropy due to the increase of the number of interfaces \cite{nie_magnetization_2010,nemoto_analysis_2005}. The \vmcd signal is approximately one order of magnitude larger than the Kerr signal, regardless the number of repetitions. We stress out the fact that the signal to noise ratio is also significantly larger when using the \vmcd technique, the lock-in detection parameters (filtering and averaging) being the same for both techniques. We also observe that \vmcd increases when decreasing the number of repetitions whereas the Kerr effect signal decreases as shown in \fig{7}{c} and \blue{d}. It confirms the fact that this technique is very interesting to detect the magnetization of ultra-thin ferromagnets where the Kerr effect signal is barely detectable using a conventional Si-based photodiode. 


In order to better understand the thickness dependence of the MCD signal, we consider $\lambda_L$ and $\lambda_R$, the absorption length of the \copt film for left and right circular helicities. We define the average absorption length as:$\lambda = (\lambda_L + \lambda_R)/2$ and the contrast of absorption due to the MCD as: $\delta = (\lambda_L - \lambda_R)/2 $. The transmitted light intensity is expressed as $I_{R(L)} = I_0 .\exp \left( -t/\lambda_{R(L)} \right)$ for the right-handed (left-handed) circularly-polarized light.
The Seebeck voltage is given by the temperature difference which is proportional to the light intensity: $V_{\textup{Seebeck}} = S \Delta T = A.S.I$, where $A$ is a constant of the material.
If we now assume that $\delta << \lambda$, we obtain the following expression for \vmcd :

\begin{equation}
V_{\textup{MCD}} = V_L - V_R = A.S.(I_L - I_R) =A.I_0.S. \frac{\delta .t}{\lambda^2} \exp \frac{-t}{\lambda}
\end{equation}

This relation shows that unlike the Kerr effect, the MCD signal has an optimum of sensitivity when $t=\lambda$. In our case, we can see that the \vmcd signal is already in the exponential decrease regime, implying that the optimum of sensitivity is below $n = 2$ (equivalent to 4.6 nm). This also indicates that the techniques will be most suited for ferromagnetic metals, where the absorption length is in the nanometer range. 
	
\begin{figure}[h!]
\begin{center}
\includegraphics[width=0.5\textwidth]{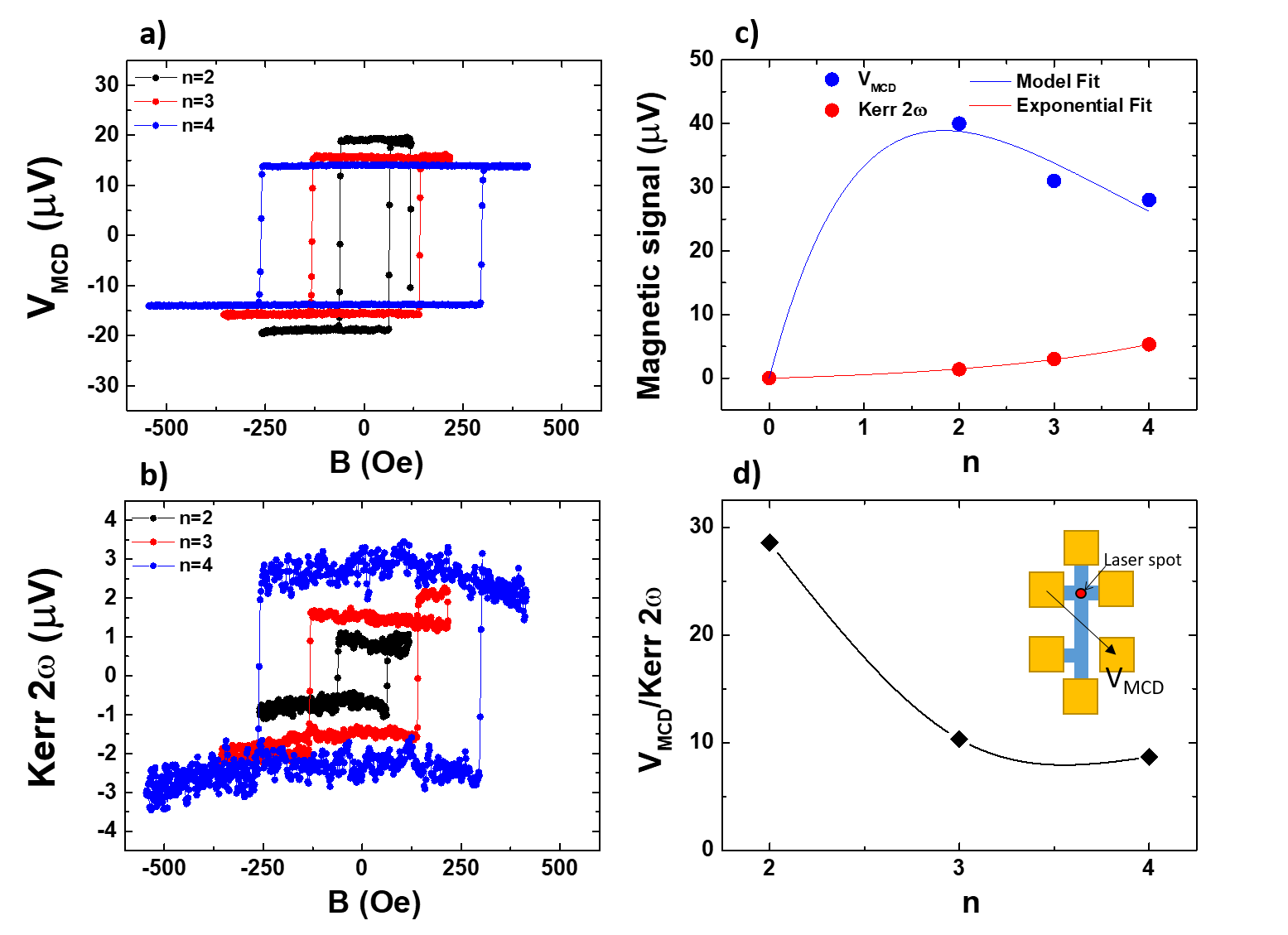} 
\caption{(color online) Hysteresis loops for different (Co/Pt) repetitions ($n = 2,3,4$) simultaneously measured by a) the \vmcd technique and b) the Kerr effect. The beam is focused on the center of the Hall bar for each \copt sample and the laser power is 650 \textmu W. c) Summary of the two magnetic signal dependence with the Co/Pt repetition number. d) Ratio between the \vmcd and the Kerr effect signals as a function of the number of repetitions.}
\label{Fig7}
\end{center}
\end{figure}

	
In order to further confirm the nature of the \vmcd signal, we vary the incident light polarization. The PEM is used to control the light helicity, the retardation is fixed at 0.25 $\lambda$. As shown in \fig{8}{a}, we rotate the entrance polarizer with respect to the PEM axes. The rotation angle is defined as $\alpha$ (see \fig{8}{a} and \blue{b}). The light polarization is linear when the entrance polarizer and the PEM axis are aligned ($\alpha =~0$\degree , 90 \degree , 180 \degree and 270 \degree). The light polarization is circular for $\alpha =~45$\degree and 225\degree (right handed), and  for $\alpha =~135$\degree and 315\degree (left handed).

\begin{figure}[h!]
\begin{center}
\includegraphics[width=0.5\textwidth]{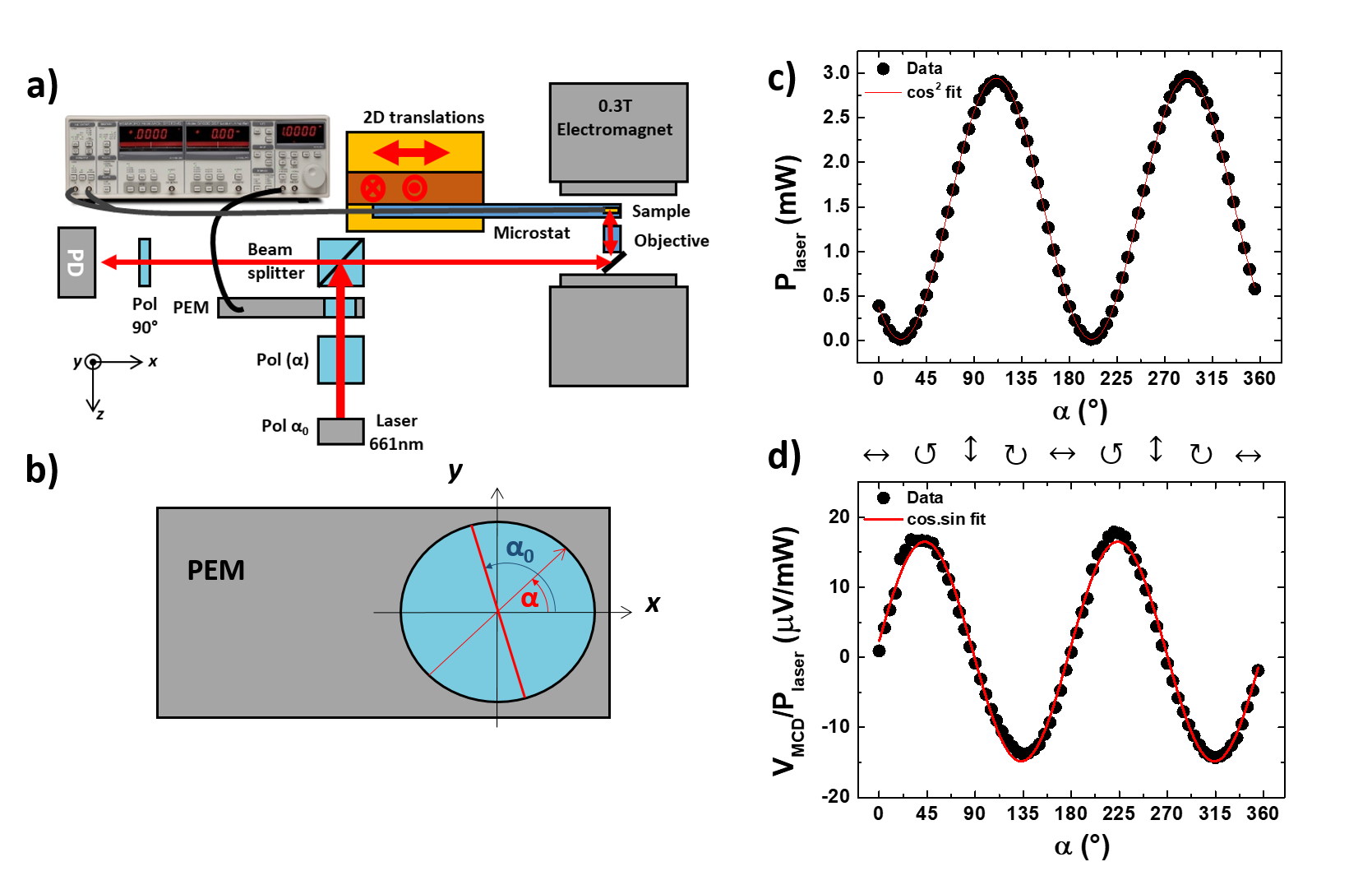} 
\caption{(color online) a) Schematic top view of the scanning confocal setup, the light polarization is obtained by associating a linear polarizer and a photoeleastic modulator. b) Definition of the angle $\alpha$, $x$ and $y$ denote the PEM optical axis. c) Laser power dependence with the polarizer angle $\alpha$. d) \vmcd signal normalized by the laser power, the top insets indicate the light polarization states.}
\label{Fig8}
\end{center}
\end{figure}
 
Here, we focus on the \coptttt. A 1000 Oe external magnetic field is first applied to saturate the magnetization along the $+z$ direction, it is then turned to zero to measure the remanent magnetization state. The laser beam is focused on the Hall bar center and the dependence on the polarizer angle is recorded.
The laser beam is already polarized out of the optical fiber, so the transmitted laser power is also affected by the polarizer rotation. The laser power is also recorded using a powermeter for each polarizer angle. As shown in \fig{8}{c}, $P_{laser}$  follows the Malus law: 

\begin{equation}
P_{laser}=P_0\cos^2\left(\alpha-\alpha_0\right)
\end{equation}

Where $P_0$ is the nominal laser power, $\alpha$ is the angle between the polarizer and the PEM optical axis and $\alpha_0$ is the angle between the initial laser beam polarization and the first polarizer (see \fig{8}{b}). The power dependence on the polarizer angle gives minima for $\alpha =~20 $\degree  and 200\degree, indicating that $\alpha_0 =~110$\degree . In order to correctly measure the \vmcd signal, we have to normalize the recorded \vmcd by $P_{laser}$. 
The dependence of \vmcd on the polarizer angle is reported in the \fig{8}{d}, the inset on top shows the incident light polarization state. We observe a $\cos\alpha\sin\alpha$ angular dependence: \vmcd vanishes when $\alpha =~$0\degree , 90\degree , 180\degree and 270\degree, i.e. when the light polarization is linear. It shows minima (maxima) for $\alpha =~$45\degree and 225\degree ($\alpha =~$135\degree and 315\degree) for $\sigma_+$ and $\sigma_-$ light polarizations respectively. This result emphasizes the fact that the detected voltage is due to the different absorption of circularly polarized light by the ferromagnetic film, resulting in different photovoltages in Ge  for clockwise and counterclockwise light helicities. 

\begin{figure}[h!]
\begin{center}
\includegraphics[width=0.4\textwidth]{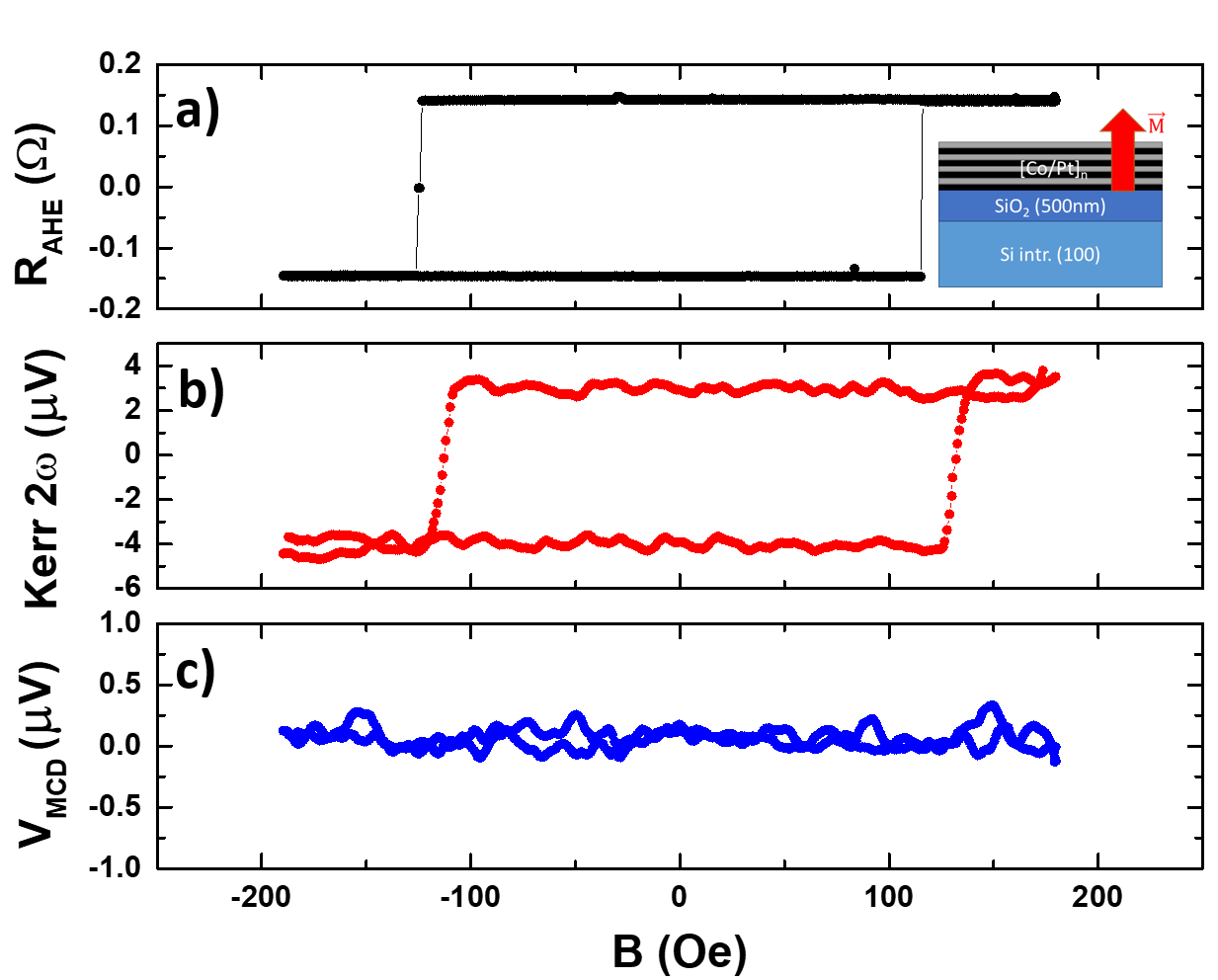} 
\caption{(color online) a) AHE hysteresis loop recorded with a current $I_{\textup{DC}} = 100$\uA . b) MOKE hysteresis loop using a 100 \% circularly polarized red light ($\lambda = 661$~nm) focused on the Hall bar center, the spot size is about 1.5\um. c) \vmcd hysteresis loop, the voltage is demodulated at the PEM frequency $f$ and measured between two Hall bar contacts.}
\label{Fig9}
\end{center}
\end{figure}

Finally, to prove that the \vmcd signal is related to a photovoltage generated in Ge and not directly in the ferromagnetic film, we have grown a \coptt film on a SiO$_2$ substrate and patterned the same Hall bars. Again, the magnetic properties are measured using simultaneously the AHE, \vmcd and the Kerr effect. As shown in \figsimple{9}, the magnetic anisotropy is also perpendicular, the hysteresis loop can be detected using the AHE or the Kerr effect but there is no \vmcd signal. This result confirms that the measured photovoltage comes from Ge due to MCD in the (Co/Pt) film.
	
	\section{Application to the study of magnetic domain wall motion}	
	
We exploit the MCD detection technique to image multidomain magnetic configurations and the motion of domain walls. Here, we focus on the \copttt sample. We introduce a magnetic domain wall in the Hall bar by applying a specific magnetic field sequence and use the two magnetic microscopy techniques (Kerr effect and MCD detection) to image the domain wall propagation. We repeat the following field sequence: the magnetization is first saturated along $+z$, then a negative magnetic field $B_{\textup{nucl}}$ is applied to nucleate domains and we image the magnetic configurations. The sequence is iterated by increasing $\lvert B_{\textup{nucl}}\lvert$ in order to move the domain wall.

\begin{figure}[h!]
\begin{center}
\includegraphics[width=0.5\textwidth]{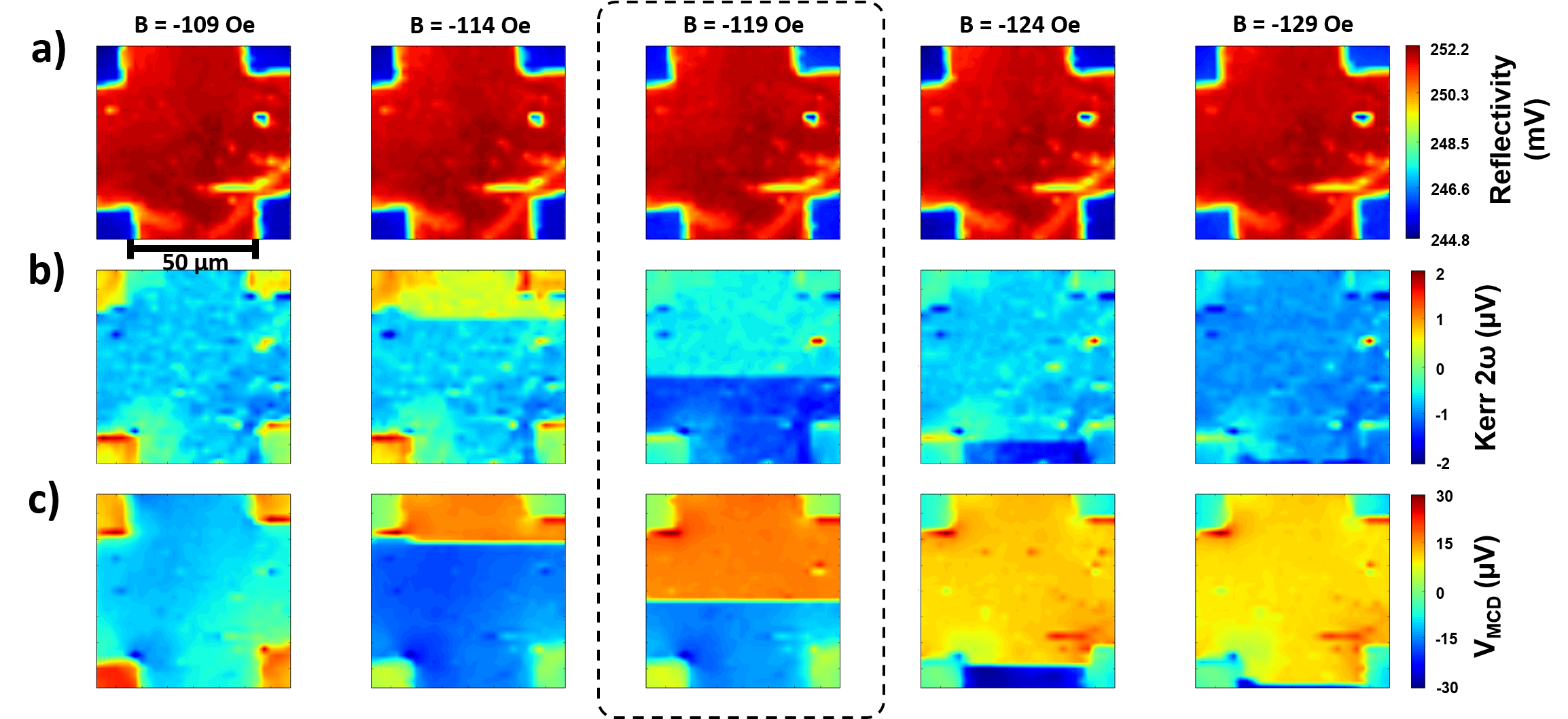} 
\caption{(color online) a) Reflectivity maps. b) \vmcd maps. c) Kerr effect maps for different applied magnetic fields. Before each two-dimensional scan, a +500 Oe field is first applied to saturate the magnetization along the $+z$ direction, a precise negative magnetic field value is then applied to nucleate and propagate a domain wall. We can see the domain wall propagating when increasing the magnitude of the magnetic field.}
\label{Fig10}
\end{center}
\end{figure}

The magnetic configuration can be imaged simultaneously using the Kerr effect microscopy and the electrical detection of the local magnetization based on the MCD in the (Co/Pt) film. \fig{10}{a} shows the reflectivity of the sample for different magnetic field intensities, the (Co/Pt) (resp. Ge) film corresponds to the red (resp. blue) color. \fig{10}{b} and \blue{c} show the Kerr effect and \vmcd maps recorded for the different applied magnetic fields. For $B = -109$~Oe, the magnetization is still saturated and uniform on the Hall bar scanned area. Then, by iterating the magnetic field sequence, we see a domain wall propaating in the Hall cross, the magnetic domains are pointing toward $+z$ (in red) and $-z$ (in blue). For $B = -119$~Oe, corresponding to the box delimited by a black dashed line in \fig{10}, the domain wall is located in the middle of the Hall cross. By further increasing the negative magnetic field, we observe the propagation of the wall along the $-y$ direction. Interestingly, the domain wall (resp. its propagation) is perpendicular (resp. parallel) to the current applied in the Hall bar.

Here, by imaging the two-dimensional magnetization maps simultaneously with the two techniques, we conclude about the very high sensitivity of the MCD detection technique. 


\section{Conclusion}

To conclude, we have successfully grown perpendicularly magnetized thin films on a Ge (111) substrate. The magnetic properties of (Co/Pt) multilayers were investigated using the anomalous Hall effect, magneto-optical Kerr effect microscopy and a new hybrid electro-optical technique based on the magnetic circular dichroism in (Co/Pt) and combining the thermoelectric and semiconducting properties of Ge. Our study reveals that this hybrid technique shows several advantages for the magnetic characterization of ultra-thin films and could be generalized to a large variety of semiconducting (Si, GaAs...) and thermoelectric substrates (Bi, Bi$_2$Se$_3$ ...).

The detection being electrical, it is particularly well suited for future investigation of the magnetic properties of 2D materials, where the standard magnetic imaging techniques are difficult to setup. Moreover, both the signal and the signal-to-noise ratio are much larger than the Kerr effect ones.

We showed that the electrical detection of the magnetic circular dichroism of (Co/Pt) originates from the Seebeck effect, as a result of the difference of thermal gradients between the two electrical contacts. We demonstrated that the measurement geometry can be optimized in order to maximize this thermal contribution and obtain a uniform measurement by using strongly asymmetric contacts. 
Alternatively, by applying a bias current parallel to the detection axis, one can suppress or enhance the total sensitivity of the technique by combining the thermal and drift contributions. The drift component can also be isolated by using its symmetries with respect to the bias current in order to obtain a measurement almost independent of the geometry. \newline

Finally, we point out the fact that it is not necessary to connect electrically the ferromagnetic film, the two contacts can simply be made on the semiconducting substrate close to the ferromagnet. This feature added to the high sensitivity of the technique in the ultra-thin film regime makes this technique an excellent alternative to traditional magnetometry for the investigation of ferromagnetism in the emergent 2D ferromagnets grown (or transfered) on semiconductors.

\section{Acknowledgements}

The authors acknowledge the financial support from the ANR projects ANR-16-CE24-0017 TOP RISE and ANR-18-CE24-0007 MAGICVALLEY as well as Dr. Carlo Zucchetti and Dr. Federico Bottegoni from the Politecnico di Milano (Italy) for setting up the microscope and for fruitful discussions.

\end{sloppypar}

\end{document}